\documentclass[showpacs,amsmath,amssymb,twocolumn,superscriptaddress,prl]{revtex4}
\usepackage{times}
\usepackage{graphicx}
\usepackage{amsfonts}
\usepackage{amsmath, amsthm, amssymb}
\usepackage{dsfont}
\usepackage{subfigure}
\usepackage{epstopdf}

\renewcommand{\section}[1]{{\par\it #1. }\ignorespaces}

\def\i{\mathrm{i}}

\begin{document}
\title{Magnetization dynamics and Majorana fermions in ferromagnetic Josephson junctions along the quantum spin Hall edge}
\author{Annica M. Black-Schaffer}
 \affiliation{NORDITA, Roslagstullsbacken 23, SE-106 91 Stockholm, Sweden}
 \author{Jacob Linder}
\affiliation{Department of Physics, Norwegian University of Science and Technology, N-7491 Trondheim, Norway}
\date{Received \today} 
\begin{abstract}
We investigate the interplay between ferromagnetic and superconducting order at the edge of a quantum spin Hall insulator. Using complementary analytical and self-consistent numerical approaches, we study a ferromagnetic Josephson junction and show how the direct coupling between magnetism and the superconducting $U(1)$ phase gives rise to several unusual phenomena which distinguishes the present system from its non-topological equivalent. In particular, we demonstrate how the anomalous current-phase relation triggers supercurrent-induced magnetization dynamics and also study the spatial localization of the Majorana fermions appearing in the junction.
\end{abstract}

\maketitle


The class of materials known as topological insulators (TIs), or quantum spin Hall insulators (QSHIs) in two dimensions, represents a new quantum-state of matter \cite{Hasan10}. Its hallmark is the appearance of topologically protected spin-polarized edge-states which are robust against time-reversal symmetry invariant perturbations. Recently, it has been realized \cite{Fu08,Fu09} that such systems can host Majorana fermions which obey non-Abelian statistics. Thus, the study of TIs is presently generating much interest, both from a fundamental physical point of view \cite{Wilczek09} and in terms of future applications in fault-tolerant topological quantum computation \cite{Nayak08, Akhmerov09,Fu09b}.
An intriguing question relates to how different types of long-range orders may accommodate to the unusual electronic environment in a TI. In particular, several works have recently focused on how proximity-induced superconductivity is manifested in TIs \cite{Fu08, Fu09, Linder10, Stanescu10, Black-Schaffer10QSHI}. By adding ferromagnetic correlations via a proximate material with the desired magnetism, it has been realized that it is possible to generate Majorana fermions in such TI hybrid structures \cite{Fu08, Fu09}. Moreover, it was predicted that an anomalous current-phase relation (CPR) would appear for a supercurrent flowing through a topological Josephson junction \cite{Tanaka09}. Interestingly, it follows from this fact that a direct coupling between the magnetic and superconducting order exists which can be manipulated simply by current-biasing the system. 

\begin{figure}[htb]
\includegraphics[scale = 0.25]{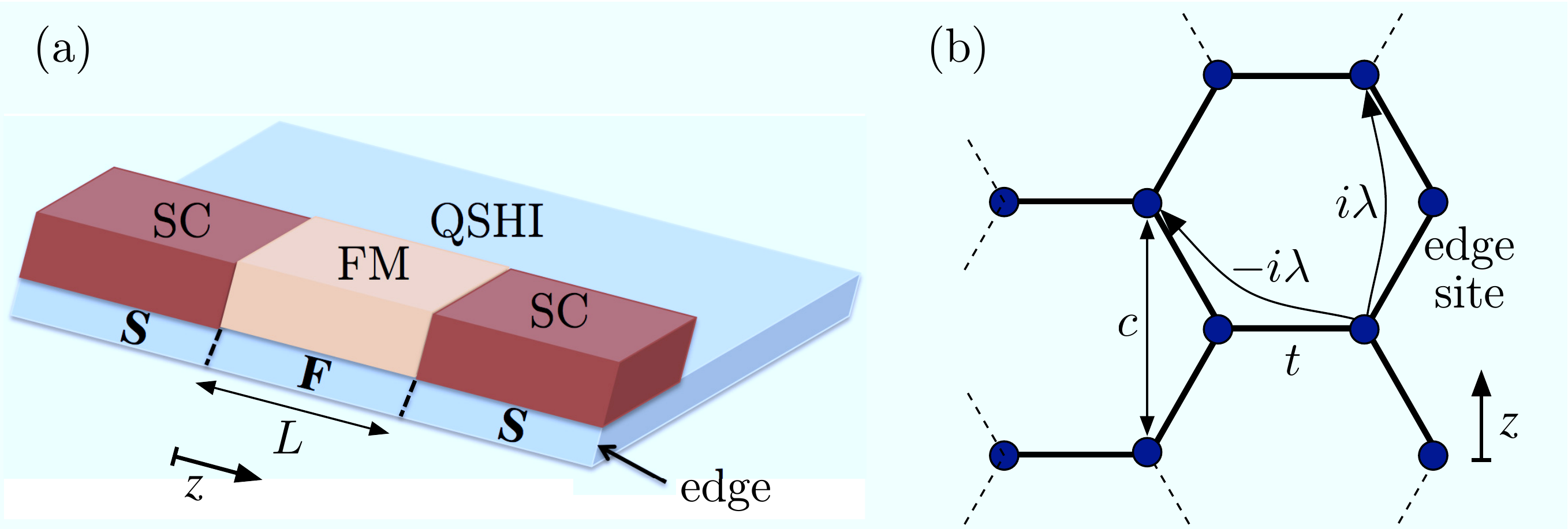}
\caption{\label{fig:setup} (Color online) (a): Schematic of proposed experimental setup. Two superconducting contacts (SC) induces superconductivity in the S regions of the QSHI whereas the ferromagnetic insulator (FM) induces an effective magnetic field in the F region. (b): Kane-Mele model for the QSHI with a zigzag edge. For details, see text. 
}
\end{figure}

In this Rapid Communication, we address a series of unusual phenomena that arise at the edge of a QSHI due to the interplay between ferromagnetism and superconductivity that have no counterpart in a non-topological equivalent of such a system. We employ both an analytical continuum and a self-consistent microscopic numerical model, the latter accounting for non-ideal effects such as the partial depletion of the superconducting order parameter near the interface regions. First, we demonstrate that the predicted anomalous CPR in a superconductor$\mid$ferromagnet$\mid$superconductor (S$\mid$F$\mid$S) QSHI Josephson junction, see Fig.~\ref{fig:setup}(a), is robust towards non-ideal effects in the system, but that it still features different behavior compared to the prediction of analytical models. Second, we demonstrate that this anomalous CPR can be experimentally observed and manipulated by means of supercurrent-induced magnetization dynamics. This is done by numerically solving the full Landau-Lifshitz-Gilbert equation to obtain the time-dependence of the magnetization under the influence of an AC Josephson effect. Finally, we study the localization of Majorana fermions appearing at the S$\mid$F interface regions, a property crucial with regards to actual experimental schemes intending to exploit their non-Abelian properties. 
%

For the self-consistent microscopic treatment of a QSHI S$\mid$F$\mid$S junction we consider a minimal microscopic model $H = H_0 + H_S + H_F$ defined on the honeycomb lattice. Here
$H_0  =  -t \sum_{\langle i,j\rangle,\alpha} a_{i\alpha}^\dagger a_{j\alpha} + \mu_{i} \sum_{i,\alpha} a_{i\alpha}^\dagger a_{i\alpha} +
i \lambda \sum_{\langle \langle i,j\rangle \rangle} \nu_{ij}a_{i\alpha}^\dagger \sigma^z_{\alpha \beta}a_{j\beta}$
%
is the Kane-Mele model for a QSHI \cite{Kane05} with $a$ the fermion operator, $\alpha, \beta$ the spin indices, ${\boldsymbol \sigma}$ the Pauli matrices,$\langle i,j \rangle$ and $\langle \langle i,j \rangle \rangle$ denoting nearest neighbors and next nearest neighbors respectively, and $\nu_{ij} = +1$ $(-1)$ if the electron makes a left (right) turn to get to the second bond. $t=1$ is the nearest neighbor hopping amplitude and we also fix the spin-orbit coupling $\lambda = 0.3$ which gives a bulk band gap of 1. Using a unit cell length of $c = 1$ we get a linear dispersion coefficient $\hbar v_F = 1.15$ for the Dirac-like edge state which we choose to be located along the zigzag edge, see Fig.~\ref{fig:setup}(b) for details. We set the chemical potential $\mu_{i}= \mu_S = 0.3$ in the S regions, assumed large due to induced doping from the external superconducting contacts, whereas we have studied F regions with  $\mu_F = 0, 0.1$. 
%
%
The superconductivity in the S regions of the QSHI is modeled using an attractive Hubbard-$U$ pairing potential which is induced by the external superconducting contacts \cite{Black-Schaffer10QSHI} such that 
$H_S  =  \sum_i U_i a_{i\uparrow}^\dagger a_{i\uparrow} a_{i\downarrow}^\dagger a_{i\downarrow},$
%
where $U_i = U$ in the S regions but zero otherwise. Treating $H_S$ in mean-field theory, the self-consistency condition for the superconducting order parameter $\Delta$ reads
$\Delta_i  =  -U_i \langle a_{i\downarrow} a_{i\uparrow} \rangle$.
%
In general, we choose $U$ such that the superconducting coherence length $\xi = \hbar v_F/\Delta_0$ in the interior of the S regions along the edge satisfies $\xi \sim L$, where $L$ is the junction length.
%
Finally, $H_F$ models the influence of the external ferromagnet insulator on the QSHI by 
$H_F  =  \sum_{i }  a_{i\alpha}^\dagger {\boldsymbol h}_i\cdot {\boldsymbol \sigma}_{\alpha \beta} a_{i\beta},$
%
where ${\boldsymbol h}_i = (h_x,h_y,h_z)$ is the (constant) induced magnetic field in the F region of the QSHI, ignoring any orbital contribution.
%
Further, we let both the S and F regions extend well into the bulk of the QSHI and we assume sharp S$\mid$F interfaces. We solve the above model self-consistently by first diagonalizing the Hamiltonian with the order parameter $\Delta_i = \Delta_0$ in the S regions. $\Delta_i$ is then recalculated using the self-consistency condition. By repeating this process until $\Delta_i$ does not change between two subsequent iterations, we achieve self-consistency for the superconducting state and thus capture the inverse proximity effect (IPE), i.e. the decrease of $\Delta$, or equivalently loss of Cooper pairs, on the S side of the junction. 
By applying a difference $\Delta \phi$  in the $U(1)$ phase between the order parameters in the two S regions we can also calculate the Josephson supercurrent $I$ through the junction using the continuity equation for the charge current \cite{Black-Schaffer08}. However, the actual phase drop $\phi$ across the junction itself, i.e. across F, necessarily satisfies $\phi \leq \Delta \phi$ since a finite current will always cause a finite phase drop even in the S regions. We ensure full self-consistency by only fixing the phase in the outermost regions of the S regions. For further details on the self-consistency method see Refs.~\cite{Black-Schaffer08, Linder10SFS}.

Our non-self-consistent treatment of a QSHI S$\mid$F$\mid$S junction relies on fixing $\Delta = \Delta_0$ in the S regions and using the continuum field-theoretic model for the QSHI edge state introduced by Fu and Kane \cite{Fu09}. The corresponding Hamiltonian is written as: 
\begin{align}
H = \begin{pmatrix}
\boldsymbol{\sigma}\cdot(v_F\boldsymbol{k} + \boldsymbol{h}) & \i\sigma_y\Delta \\
-(\i\sigma_y\Delta)^* & \boldsymbol{\sigma}^*\cdot(v_F\boldsymbol{k} - \boldsymbol{h})\\
\end{pmatrix}
\end{align}
According to the geometry in Fig. \ref{fig:setup}, $\boldsymbol{k} = k_z\hat{\boldsymbol{z}}$. For an arbitrary magnetization direction in the F region, an analytical expression for the Andreev-bound states in a short, $L<\xi$, S$\mid$F$\mid$S junction can be obtained by constructing the scattering eigenstates and matching them at each S$\mid$F interface. Performing this calculation along the QSHI edge, we obtain the bound-state energies $\varepsilon = \pm \varepsilon_0$, where $\varepsilon_0 = \Delta_0\sqrt{D}\cos(\phi/2 - h_zL)$
and $D = [1+\sinh^2(kL)]^{-1}$ and $k=(h_x^2+h_y^2)^{1/2}$. When $h_y=h_z=0$, this reduces to the result of Ref. \cite{Fu09}. Interestingly, the $z$-component of the exchange field (parallel with the junction) effectively renormalizes the superconducting phase-difference \cite{Tanaka09}. This occurs due to the linear Dirac energy-momentum dispersion of the normal-state Hamiltonian for the topological edge-state, which effectively makes the exchange field enter as a vector potential. The periodic dependence of the current on the exchange field is reminiscent from the physics of S$\mid$F$\mid$S systems in conventional metals, with the important distinction that these do not depend on the magnetization orientation \cite{rmp_SFS}. We shall later demonstrate that this fact can be both experimentally observed and exploited for practical purposes via magnetization dynamics induced by a current bias along the edge of the topological insulator. Prior to this we will report on the consequences of self-consistency for the CPR.

%
%
\section{CPR}
The CPR is one of the most important properties of a Josephson junction. The non-self-consistent continuum model result may be obtained from the above expression for the Andreev-bound state $\varepsilon$, and gives that $I \propto \Delta_0 \sin(\phi/2-h_z L){\rm sgn}[\cos(\phi/2-h_z L)]$ as $T\to0$, i.e. the maximum current is reached at the critical phase $\phi_c \to \pi$ for fields along the $x$ or $y$-direction, whereas fields along the $z$-direction shifts this value. This shift is precisely due to the fact that the exchange field enters as a vector potential and thus gives rise to an effective flux along the edge. 
We note in passing that in the non-thermalized regime, it is possible to obtain a 4$\pi$-periodic CPR on a topological surface \cite{Fu09}.

\begin{figure}[htb]
\includegraphics[scale = 0.8]{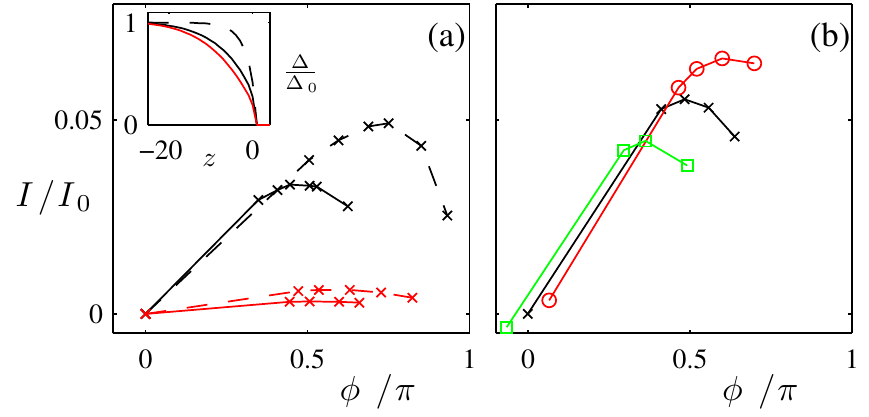}
\caption{\label{fig:CPR} (Color online) The supercurrent $I/I_0$ with $I_0 = e\Delta_0/\hbar$ and $\xi = 13$ as function of phase drop $\phi$ across the junction for (a): $h_x = 0.3$, $\mu_F = 0.1$, and $L =12$ (black) and $L= 24$ (red) with dashed lines showing results after one self-consistency step only, and (b): $h_z = 0$ (black,$\times$), 0.01 (red,$\circ$), -0.01 (green,$\square$), $\mu_F = 0$, and $L =12$. Inset shows the IPE at the S$\mid$F interface by displaying $\Delta/\Delta_0$ for $L = 12$, $\xi = 13$ (black), $\xi = 5$ (dashed) and $L = 24$, $\xi = 13$ (red).
}
\end{figure}

The self-consistent microscopic results for non-zero $h_x$ and $h_z$ is shown Figs.~\ref{fig:CPR}(a) and (b), respectively. 
%
%
A finite $h_x$-field causes Majorana modes to appear at the S$\mid$F interfaces \cite{Fu09} and we show in Fig.~\ref{fig:CPR}(a) results for both $L = 0.9 \xi$ and $L = 1.8\xi$. We clearly see that $\phi_c$ is shifted down toward $\pi/2$ for both cases. 
We attribute this shift to a significant amount of IPE, as seen in the inset. In fact, processes influencing the S regions are known to shift $\phi_c$ below the standard value of $\pi/2$ in conventional S$\mid$N$\mid$S junctions \cite{Golubov04}. Here we see a similar shift but starting from the non-self-consistent solution at $\phi_c \to \pi$. To support this conclusion we show in the dashed lines the CPR after the first step of the self-consistency loop, starting from the analytical $\Delta = \Delta_0$  step-function solution. Already after one iteration we see that $\phi_c$ is significantly reduced while at the same time a sizable IPE has appeared.
We can quantify the IPE by fitting $\Delta$ close to the interface to the form $\Delta = \Delta_0 [1 - \exp(-z/\kappa_{\rm IPE})$]. As expected, the decay length $\kappa_{\rm IPE}$ is linearly proportional to $\xi$, although the linear coefficient is somewhat dependent on both $L$ and $h_x$, but not on $\mu_F$. Thus, by varying $\xi$ we can significantly modify the IPE and for very short $\xi$ we indeed see $\phi_c$ increasing, but never fully to $\pi$. 
On the other hand, we only achieve $\phi_c<\pi/2$ due to current depairing \cite{Black-Schaffer10Tdep}, which is only present in artificially short junctions. Therefore, for any experimental junction, where $\xi$ is significantly larger than we can model, we predict $\phi_c \approx \pi/2$.

Figure \ref{fig:CPR}(b) shows the results for finite $h_z$-fields. There are no Majorana fermions appearing for this field direction, but still, a $h_z$-field has interesting consequences for the CPR. As predicted analytically, we see a linear shift in $\phi_c$ with $h_z$ and $L$, however, the linear coefficient is $\sim1.5$. Moreover, we again see $\phi_c \approx \pi/2$, and not $\pi$, in the limit of zero field. The self-consistent CPR with a $h_z$-field can thus, up to numerical accuracy, be written as $I = I_c \sin(\phi - 1.5h_z L)$. We also see a shift in $I_c$ with $h_z$ (but not $L$), such that $I_c = I_c(h_z = 0) + 0.1h_z$. Such a dependence is not seen in the analytical solution but can be traced back to a changed IPE. In fact, when the IPE increases, we see a concurrent drop of both $I_c$ and $\phi_c$, which can also explain the larger linear coefficient for the field-induced phase shift than found analytically.

%
\section{Magnetization dynamics}
The direct coupling between the superconducting phase difference and the exchange-field offers an interesting opportunity: \textit{supercurrent-induced magnetization dynamics}. By biasing the system with a supercurrent, one would expect to see a time-evolution of the magnetic order parameter. To address this issue, we make use of the Landau-Lifshitz-Gilbert (LLG) equation \cite{llg} (the modification of the LLG equation on a topological surface in the presence of a single-particle charge-current was recently discussed in Ref. \cite{Yokoyama10}):  $\partial_t\boldsymbol{M} = -\gamma \boldsymbol{M} \times \boldsymbol{H}_\text{eff} + (\alpha/M_0) (\boldsymbol{M} \times \partial_t \boldsymbol{M}),$
where $\boldsymbol{M}$ is the magnetization vector, $M_0=|\boldsymbol{M}|$, $\gamma$ is the gyromagnetic ratio, and $\alpha$ is the Gilbert damping constant. The components of the effective field are obtained from the free energy $\mathcal{F}$ of the system via: $(\boldsymbol{H}_\text{eff})_j = -\frac{1}{\mathcal{V}}\frac{\partial \mathcal{F}}{\partial \boldsymbol{M}_j}$, where $\mathcal{V}$ is the volume of the F region. In what follows, we investigate how an equilibrium supercurrent can induce strong magnetization dynamics in the present system. Thus, we focus on the effect of the Andreev-bound states, the rationale being that it is precisely this contribution that constitutes the Josephson-current induced magnetization dynamics, and assume that other contributions such as anisotropy fields can be neglected. We underline the fact that the analytical expression for the bound-states captures the field-dependence exhibited by the effective superconducting phase difference, as also confirmed by our self-consistent results, which is the key to the supercurrent-induced magnetization dynamics. Obtaining the free energy from $\mathcal{F} = -\frac{2}{\beta} \text{ln}[2\cosh(\beta\varepsilon_0/2)]$, one may rewrite the LLG equation in the convenient dimensionless form: $\partial_\tau\boldsymbol{n} = -\boldsymbol{n}\times\boldsymbol{\mathcal{H}}_\text{eff} + \alpha(\boldsymbol{n}\times\partial_\tau \boldsymbol{n})$, where we have defined $\boldsymbol{n} = (n_x,n_y,n_z)=\boldsymbol{M}/|\boldsymbol{M}|$, $\tau=\omega_Jt$ where $\omega_J$ is the Josephson frequency, and finally: 
$(\boldsymbol{\mathcal{H}}_\text{eff})_j = C_1n_j\tanh(\beta\epsilon_0/2)\cos(\phi/2-hLn_z),\; j=x,y$, $
(\boldsymbol{\mathcal{H}}_\text{eff})_z = C_2\tanh(\beta\epsilon_0/2)\sin(\phi/2-hLn_z).
$
Here, $C_1 = -\Delta_0h^2L\sinh(kL)/(2kD^{5/2}\omega_J\Gamma)$ and $C_2=\Delta_0\sqrt{D} hL/(2\omega_J\Gamma)$ where $\Gamma = M_0\mathcal{V}/(2\gamma)$. The inverse temperature is $\beta$, and we set $T/T_c=0.2$.
%
\begin{figure}[t!]
\centering
\resizebox{0.42\textwidth}{!}{
\includegraphics{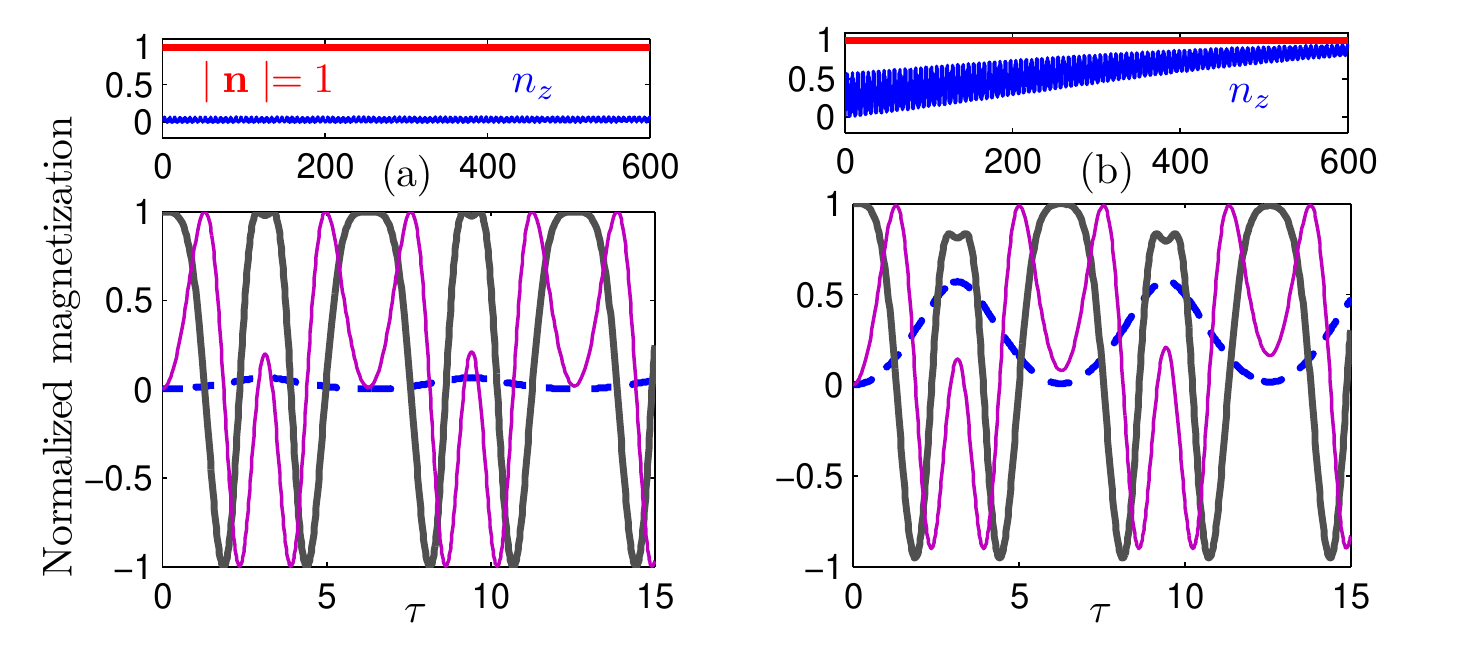}}
\caption{(Color online) Time-evolution of the magnetization components for (a) $\alpha=0.01$ and (b) $\alpha=0.1$. Thick dashed line: $n_z$, thick full line: $n_y$, thin full line: $n_x$. The top panels show the magnetization dynamics on a larger time-scale.}
\label{fig:magdyn} 
\end{figure}

We have solved the above LLG equation numerically to simulate the magnetization dynamics induced by a Josephson current. It should be noted that a Josephson current does not induce magnetization dynamics in a conventional S$\mid$F$\mid$S junction, whereas this feature appears in the present context due to the Dirac-like dispersion with momentum-spin locking for the topological edge states. Let us estimate the magnitude of the parameters entering in the expressions for $C_1$ and $C_2$. The proximity induced superconducting gap is taken to be small, $\Delta_0=0.1$ meV, and we also make a conservative assumption for the exchange field, $h=5$ meV (Ref.~\footnote{It was shown in Ref.~\cite{Konig07} that a properly aligned magnetic field $B$ would induce an exchange energy $h = B \times (3.1$ meV/T) in the edge-states.}). The length $L$ of the junction can be varied experimentally, but is here fixed to $L=400$ nm. We use a typical value for the Josephson frequency, $\omega_J = 1$ GHz. The parameter $\alpha$ is phenomenological, and we will contrast the small damping scenario $\alpha\ll1$ with strong damping $\alpha \sim 0.1$. We also treat $\Gamma$ as a free parameter to vary, since the exact values of $M_0$ and $\mathcal{V}$ will differ depending on the specific materials and dimensions chosen. When $\Gamma\gg1$, the effective field in the junction is reduced and the magnetization dynamics vanishes. Therefore, we here set $\Gamma=0.2$ and compare in Fig. \ref{fig:magdyn} the case of weak ($\alpha=0.01$) and strong ($\alpha=0.1$) damping. The initial condition is chosen to have the magnetization $\parallel \hat{\boldsymbol{y}}$, i.e. in the plane of the topological insulator. In the weak damping regime, the magnetization oscillates unabatedly as $\tau$ increases (only the main periodic pattern is shown in Fig. \ref{fig:magdyn} for clarity). For larger damping, the dynamics becomes more interesting. In this case, it is seen how the $z$-component becomes dominant and how the oscillation of the other components slowly decay as $\tau$ increases. This can be understood from the fact that the magnetization will saturate on a shorter time-scale when $\alpha\ll1$ is no longer satisfied. In fact, the right panel of Fig. \ref{fig:magdyn} suggests that \textit{supercurrent-induced magnetization switching} might be possible: the magnetization starts out $\parallel \hat{\boldsymbol{y}}$, but is seen to converge toward $\parallel \hat{\boldsymbol{z}}$ as $\tau$ increases. An important point with respect to the presence of an exchange field in the $z$-direction is that the ground-state superconducting phase difference $\phi_0$ no longer is $\phi_0=0$ or $\phi_0=\pi$, as in conventional metallic systems. Instead, the ground-state phase is now given by $\phi_0 = 2h_zL$. This is similar to the scenario of a noncentrosymmetric ferromagnet in a Josephson junction \cite{Buzdin08}, although the origin of the $\phi_0$-state here is completely different. We also note that the same magnetization dynamics effect would be present on the surface of three-dimensional (3D) topological insulators, possibly exerting stronger dynamics in magnitude in that case due to the larger number of Andreev-bound states carrying current. Finally, it should be noted that we have assumed that the magnetic anisotropy energy is small enough to be negligible, which places restrictions on which type of materials that may be used for this purpose. This fact notwithstanding, the principle of supercurrent-induced magnetization dynamics would remain possible due to the renormalized phase-difference.

%
\section{Majorana fermions}
Our self-consistent solution also offers access to various properties of the Majorana fermions which appear at the S$\mid$F interfaces at finite $h_x$-fields. Here we concentrate on their spatial extension (i.e. the degree of localization to the S$\mid$F interface) at $\phi = 0$. This property is important for any experimental scheme aimed at exploiting their non-Abelian properties through braiding. Note that we are here not concerned with their spread into the QSHI as that is set by the width of the edge state and is therefore very narrow.
In a S$\mid$F$\mid$S junction, the two Majorana fermions appear as two in-gap, near-zero energy solutions with eigenenergies $\pm \varepsilon$ and eigenvectors $\Gamma_{\pm}$, where particle-hole symmetry dictates $\Gamma_+ = \Gamma_-^\dagger \equiv \Gamma_0$. A finite $L$ gives rise to a finite mixing between the two Majorana modes which, at $\phi = 0$, leaves a non-zero $\varepsilon$ \cite{Fu09}. Also, the two eigenvectors contain an equal mixture of the two Majorana modes living at the two separate interfaces. But, by defining the vectors $\gamma_1 = 1/2(e^{i \theta}\Gamma_0 + e^{-i \theta}\Gamma_0^\dagger)$ and $\gamma_2 = i/2(e^{i \theta}\Gamma_0 - e^{-i \theta}\Gamma_0^\dagger)$, with a variable $\theta$, we can always isolate each Majorana mode, which then also explicitly displays their $\gamma^\dagger = \gamma$ Majorana nature.
By studying the (normalized) density $|\gamma_i|^2$, we find that the spatial extent of the Majorana mode on the S side of the SF interface is only a function of $\xi$, whereas on the F side it is only a function of $h_x$. Notably, we find no significant dependence on $\mu_F$ or $L$. 
%
\begin{figure}[htb]
\includegraphics[scale = 0.8]{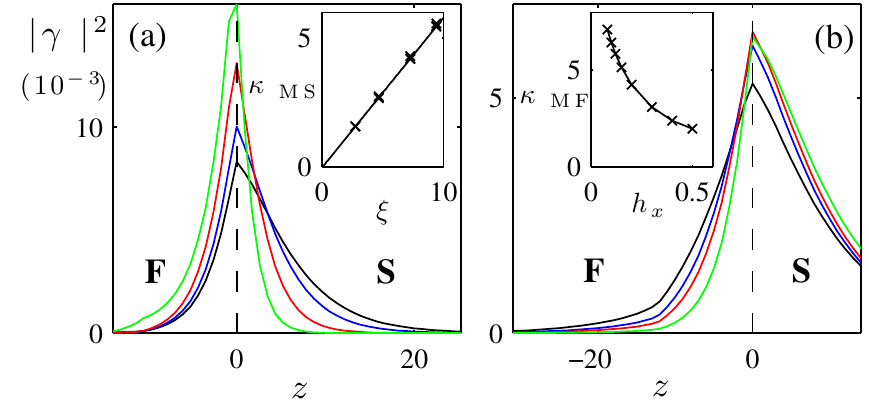}
\caption{\label{fig:Maj} (Color online) Majorana mode density (normalized to 1) along the QSHI edge as function of the distance $z$ (unit cells) from the S$\mid$F interface for $L = 12$, $\mu_F=0$, and (a): $h_x=0.3$ with $\xi = 3, 5, 7, 10$ (increasing decay length in S) and (b): $\xi = 13$ with $h_x = 0.3, 0.2, 0.15, 0.08$ (increasing decay length in F). Inset (a): Crosses show $\kappa_{\rm M,S}$ vs.~$\xi$ for $L = 8,12,24$. Line is $\kappa_{\rm M,S}=0.6\xi$. Inset (b): $\kappa_{\rm M,F}$ vs.~$h_x$. Line is only a guide to the eye.
}
\end{figure}
Figure \ref{fig:Maj}(a) shows the decay of the Majorana mode into the S region for a few representative values of $\xi$. 
Note that the slight difference in the curves on the F side for different $\xi$ is solely due to the normalization constraint, giving different peak heights at the interface.
The decay on the S side can to good accuracy be approximated as an exponential decay with a decay length $\kappa_{\rm M,S}$. We find that $\kappa_{\rm M,S}\simeq 0.6 \xi$ for all junctions we have studied (see inset). This decay length is similar to $\kappa_{\rm IPE}$ and one could speculate that the Majorana mode extends into the IPE depression region of the superconducting order parameter. However, $\kappa_{\rm IPE}$ also has a moderate dependence on $L$ and $h_x$ \footnote{E.g.~$\kappa_{\rm IPE}$ varies from $0.5\xi$ to a $0.63\xi$ when $L$ is increased from 8 to 24.} which $\kappa_{\rm M,S}$ does not. Thus the Majorana decay into the S region is not solely governed by the IPE although the two show a strong correlation.
In Fig.~\ref{fig:Maj}(b) we show a similar set of data for the decay of $|\gamma_i|^2$ on the F side of the junction for different values of $h_x$. As seen in the inset, we find a non-linear dependence on $h_x$ for $\kappa_{\rm M,F}$ (or the energy gap produced in the edge state due to $h_x$) although it is monotonic.

In summary, we have considered the interplay between ferromagnetism and superconductivity at the edge of a QSHI. We have shown that the supercurrent flowing in a QSHI ferromagnetic Josephson junction exhibits several unusual phenomena, which makes it distinct from its non-topological equivalent. Using both analytical and fully self-consistent numerical calculations, we have studied an anomalous CPR, supercurrent-induced magnetization dynamics, and the appearance of Majorana fermions along the QSHI edge. Besides the application of Majorana fermions to topological quantum computing, the interplay between magnetism and the superconducting $U(1)$ phase in this type of system opens new perspectives related to spin-polarized and tunable supercurrents.

%
\begin{acknowledgments}
A.M.B.-S. thanks Hans Hansson and Eddy Ardonne for valuable discussions.
\end{acknowledgments}


\end{document}